\documentclass[twocolumn,amsmath,amssymb,aps,nofootinbib]{revtex4}

\usepackage{amsmath}
\usepackage{amssymb}
\usepackage{amsfonts}
\usepackage{graphicx}
\usepackage{dcolumn}
\usepackage{bm}
\usepackage{natbib}

\newcommand{\beq}{\begin{equation}}
\newcommand{\eq}{\end{equation}}
\newcommand{\bea}{\begin{eqnarray}}
\newcommand{\ea}{\end{eqnarray}}
\newcommand{\p}{\partial}
\newcommand{\nn}{\nonumber}

\def\be{\begin{equation}}
\def\ee{\end{equation}}
\def\ba#1\ea{\begin{align}#1\end{align}}
\def\bg#1\eg{\begin{gather}#1\end{gather}}
\def\la{\label}
\def\er{\eqref}
\def\fr{\frac}
\def\qu{\quad}
\def\lag{\langle}
\def\rag{\rangle}

\def\({\left(}
\def\){\right)}
\def\[{\left[}
\def\]{\right]}
\def\const{\text{const}}
\begin{document}


\title{Constraints on RG Flows from Holographic Entanglement Entropy}

\author{Sera Cremonini$ ^{\,\clubsuit,\spadesuit}$}
\author{Xi Dong$ ^{\,\dagger}$}
\affiliation{$ ^\clubsuit$DAMTP, Centre for Mathematical Sciences,
University of Cambridge, Wilberforce Road, Cambridge, CB3 0WA, UK}
\affiliation{$ ^\spadesuit$George and Cynthia Mitchell Institute for Fundamental Physics and Astronomy, \\
Texas A\&M University, College Station, TX 77843, USA}
\affiliation{$ ^\dagger$Stanford Institute for Theoretical Physics, Department of Physics,
Stanford University, Stanford, CA 94305, USA}

\date{\today}

\begin{abstract}
We examine the RG flow of a candidate $c$-function, extracted from the holographic entanglement entropy of 
a strip-shaped region, for theories with broken Lorentz invariance.
We clarify the conditions on the geometry that lead to a break-down of monotonic RG flows as is expected for generic Lorentz-violating field theories.
Nevertheless we identify a set of simple criteria on the UV behavior of the geometry which guarantee a monotonic $c$-function.
Our analysis can thus be used as a guiding principle for the construction of monotonic RG trajectories, 
and can also prove useful for excluding possible IR behaviors of the theory. 
\end{abstract}

\maketitle

{\bf Introduction:}
A long-standing question in quantum field theory has been whether one can identify a suitable
function which decreases monotonically along RG trajectories from the ultraviolet (UV) to the infrared (IR).
In two dimensions the existence of such a `$c$-function' is well known \cite{Zamolodchikov:1986gt},
and its value at the fixed points of the RG flow -- where the theory becomes conformal --
matches the central charge of the corresponding CFT.
Zamolodchikov's c-theorem then leads to $c_{UV} \geq c_{IR}$,
which reflects the decrease in the effective number of degrees of freedom as one goes to lower energies.
More recently we have seen the emergence of interesting connections between $c$-theorems and the behavior
of entanglement entropy $S_{EE}$. These are particularly relevant to the realm of condensed matter physics, since
$S_{EE}$ can be an \emph{order parameter} for quantum phase transitions and topological phases -- it plays
a crucial role in describing \emph{e.g.}\ the physics of confinement/deconfinement transitions
and fractional quantum Hall systems.

For a 2D CFT, the entanglement entropy for an interval of length $\ell$ is
$ S_{EE}^{\, CFT} = \frac{c}{3}  \log\left(\frac{\ell}{\epsilon}\right) + \cdots$,
where $c$ is the central charge, $\epsilon$ a UV regulator, and we are neglecting a term independent of $\ell$.
For a QFT one can then define a $c$-function \cite{Casini:2004bw,Casini:2006es} from $S_{EE}$ via
the prescription
\beq
\label{CasiniHuerta}
c_{2D} = 3 \ell \, \frac{d S_{EE}(\ell)}{d\ell} \, ,
\eq
which is guaranteed to flow monotonically by the strong subadditivity of $S_{EE}$ combined with
Lorentz symmetry and unitarity of the underlying QFT.
Much efforts have gone into attempts to extend the $c$-theorem beyond two dimensions,
with a proof for three- and four-dimensional theories that preserve Lorentz invariance appearing only recently \cite{Casini:2012ei,Komargodski:2011vj,Komargodski:2011xv}.
AdS/CFT has allowed to recast these questions in a gravitational context,
and there is by now a well-established recipe for computing the entanglement entropy holographically
\cite{Ryu:2006bv,Ryu:2006ef} (see \emph{e.g.}\ \cite{Nishioka:2009un,Takayanagi:2012kg} for some reviews).

Given that a wide range of physical systems (notably in condensed matter physics) exhibit broken Lorentz invariance,
it is of clear interest to investigate to what extent one can recover monotonic flows in such settings.
For \emph{weakly coupled}, Lorentz-violating field theories, the fact that
entanglement does not generally decrease monotonically under RG flows was stressed recently in \cite{Swingle:2013zla}.
Here we are interested in exploring the same question but within the context of holography, as to probe
the strongly coupled regime of the underlying field theory.

We `geometrize' the question of the existence of a monotonic $c$-function -- extracted from holographic
computations of $S_{EE}$ -- by imposing null energy conditions (NEC) on the matter content of the bulk theory
which is assumed to describe Einstein gravity.
Our analysis -- while by no means exhaustive -- clarifies under what conditions
on the geometry one should find the expected breakdown of monotonicity in theories that violate Lorentz invariance.
We also identify \emph{sufficient} conditions on the asymptotic UV behavior of the geometry,
under which the $c$-function is guaranteed to decrease monotonically along the entire RG flow.
Interestingly, certain scaling solutions describing theories with a dynamical critical exponent
and hyperscaling violation obey such conditions.

{\bf Entanglement entropy and RG flows:}
We are interested in examining the behavior of a candidate $c$-function (extracted from
the entanglement entropy) in holographic models involving Einstein gravity and broken Lorentz invariance.
We assume that the field theory is $d$-dimensional, and for the purpose of computing $S_{EE}$
we consider for simplicity a strip-shaped region, whose entangling surface is described by
two parallel $(d-2)$-dimensional planes a distance $\ell$ apart.
For a $d$-dimensional CFT one then has
\beq
S^{\,CFT}_{EE} = \alpha_d \, \frac{H^{d-2}}{\epsilon^{d-2}} - \frac{1}{(d-2)\beta_d}
C_d \, \frac{H^{d-2}}{\ell^{d-2}} \, ,
\eq
where $\alpha_d,$ $\beta_d$ are numerical factors and the size of the planes $H \gg \ell$ can be thought of as an infrared regulator.
The first term encodes the standard area law, while the coefficient $C_d$ of the second term
is related to a central charge of the CFT, and can be extracted via
\beq
\label{CCFT}
C_d = \beta_d \frac{\ell^{d-1}}{H^{d-2}} \, \frac{\partial S^{\,CFT}_{EE}}{\partial \ell} \, .
\eq
In analogy with (\ref{CasiniHuerta}), this suggests a natural identification \cite{Ryu:2006ef,Myers:2012ed}
for \emph{a candidate $c$-function} along the RG flow
\beq
\label{candidate}
c_d \equiv \beta_d \frac{\ell^{d-1}}{H^{d-2}} \, \frac{\partial S_{EE}}{\partial \ell} \, ,
\eq
where $S_{EE}$ denotes the \emph{holographic} entanglement entropy and from \eqref{CCFT} we have $c_d = C_d$ at the RG
fixed points.

We would like to probe the monotonicity of the $c$-function (\ref{candidate}) for geometries that admit Lorentz violation.
To this end, we parametrize the metric by
\beq
\label{metric}
ds^2_{d+1} = - e^{\, 2 B(r)} dt^2 + dr^2 + e^{\, 2A(r)}d\vec x^2,
\eq
with the choice $A=B$ recovering Lorentz invariance.
In these coordinates, the Poincar\'e patch of AdS corresponds to $A(r) = B(r) = r/L$, with $L$ the AdS radius.
Defining
\ba
\label{fgdef}
f &\equiv  - e^{A - B } A^\prime \,,\nn\\
g &\equiv  e^{B + (d-1) A } (B^\prime - A^\prime) \,,
\ea
the NEC for
(\ref{metric}) can be written \cite{Hoyos:2010at,Liu:2012wf} as
\bea
\label{NEC1}
f^{\, \prime} = \left[ - e^{A - B } A^\prime\right]^\prime \geq 0 \, ,  \\
\label{NEC2}
g^{\, \prime} = \left[ e^{B + (d-1) A } (B^\prime - A^\prime)\right]^\prime \geq 0 \, .
\ea

The holographic entanglement entropy for a strip-shaped region is determined by minimizing
\be\la{ee}
S_{EE} = 4\pi M_p^{d-1} H^{d-2} \int_{r_m}^{r_c} dr e^{(d-2)A}\sqrt{1+ x'^2 e^{2A}}
\ee
with respect to the trajectory $x(r)$.  Here $x'\equiv dx/dr$, $r_c$ is a fixed UV cutoff, and $r_m$ is the radius of the turning point where $x'(r_m)=\infty$.  Noting that \er{ee} does not depend explicitly on $x$, we find a conserved quantity
\be
K_d \equiv e^{-(d-1)A(r_m)} = e^{-dA} \sqrt{e^{2A} + \fr{1}{x'^2}} \,,
\ee
Using this we may rewrite $S_{EE}$ and $\ell$ explicitly as functions of $r_m$:
\ba
S_{EE} &= \fr{4\pi H^{d-2}}{\ell_P^{d-1}} \int_{r_m}^{r_c} dr \fr{e^{(d-2)A}}{\sqrt{1- e^{-2(d-1)(A-A(r_m))}}} \,,\nn\\
\ell &= 2 \int_{r_m}^\infty dr \fr{e^{-A}}{\sqrt{e^{2(d-1)(A-A(r_m))}-1}} \,,\la{inrm}
\ea
where $\ell_P$ is the Planck length.

We follow the strategy of \cite{Myers:2012ed} and calculate the running
of $c_d$ as a function of $r_m$.
Making use of the chain rule $\frac{\p S_{EE}}{\p \ell} =\frac{d S_{EE}}{d r_m} / \frac{d \ell}{d r_m}$,
one then finds the simple form
\beq
\label{running}
\frac{d c_d}{d r_m} \propto - \frac{A^\prime(r_m)}{K_d}
\int_0^{\ell} dx \frac{A^{\prime\prime}}{A'^2} \,.
\eq
Here and throughout the paper we use $\propto$ to mean `proportional with positive coefficients.'  We will not repeat the derivation of \er{running} here, but refer interested readers to \cite{Myers:2012ed} for details.  We mention that all that is assumed in deriving \er{running} is that $A(r)\to\infty$ as $r\to\infty$ in the UV, as well as
\be\la{cond1}
e^{dA(r)} A'(r) \to \infty \qu \text{as} \qu r\to\infty \,.
\ee

The criterion for a monotonic $c$-function is that $d c_d/d\ell$ should have a definite sign, and in particular
it should be negative (or zero for trivial flows) in this setup.
In fact, we will recover the standard Wilsonian RG intuition provided that $d c_d/d r_m \geq 0$.
This statement relies on having $d \ell/d r_m\le 0$, which is true for minimal surfaces\footnote{Although
there may be extremal surfaces with $d \ell/d r_m> 0$,
they cannot be \emph{minimal} surfaces because we can always find another extremal surface with the same $\ell$
but a larger $r_m$, which has a smaller area and satisfies $d \ell/d r_m\le 0$.}
as argued in \cite{Myers:2012ed}, under the assumption that $A'\ge 0$ and
\be\la{cond2}
\ell \to 0 \qu \text{as} \qu r_m\to\infty \,,
\ee
where $\ell$ is determined by \er{inrm}.
Here we will always assume that the UV behavior of the geometry satisfies \er{cond1} and \er{cond2}.
The nontrivial task is then to fix the sign of $d c_d/d r_m$,
by fixing those of $A^\prime(r_m)$ and $A^{\prime\prime}$ in \er{running}.
This is the main topic of our paper.

In the Lorentz-invariant case $A=B$ studied in \cite{Myers:2012ed}, $A^{\prime\prime}$ is non-positive
according to the NEC (\ref{NEC1}).
The sign of $A'$ can be fixed if one assumes that the geometry is asymptotically AdS,
so that $A \approx r/L$ in the UV and $A'_{UV}=1/L$.  Combining this UV condition with $A''\le 0$
then guarantees that $A^{\prime}$ is everywhere positive.  It is clear that we may relax the asymptotically AdS condition as long as $A'_{UV}\ge0$.  Under this condition we then have $d c_d/d r_m \ge 0$, ensuring a monotonic $c$-function under
a Lorentz-invariant RG flow.

We are now ready to generalize the argument to the Lorentz-violating case (\ref{metric}).
Note that while the holographic entanglement entropy -- from which the candidate $c$-function \er{candidate} is defined --
is insensitive to $B(r)$, the NEC \er{NEC1} and \er{NEC2} depend nontrivially on it.
We find it useful to express (\ref{running}) in terms of $f$ and $g$,
\ba
\label{genrunning}
\frac{d c_d}{d r_m} &\propto - \frac{A^\prime(r_m)}{K_d}
\int_0^\ell dx \frac{A^{\prime\prime}}{A'^2} \nn \\
&= \frac{A^\prime(r_m)}{K_d} \int_0^\ell dx \frac{1}{A'^2}
\left[ f^\prime e^{B-A} + f \, g \, e^{-d A} \right]\,.
\ea
The second term $ \sim f \, g $ encodes the breaking of Lorentz invariance, as $g$ vanishes when the symmetry is restored.
As before, to discuss monotonicity we need to determine the sign of the integrand
\beq
\label{integrand}
f^\prime  e^{B-A} + f \, g \, e^{-d A}
\eq
as well as that of $A^\prime(r_m)$. It is useful to keep in mind
that $A^\prime(r)$ and $f(r)$ have opposite signs, as is visible from (\ref{fgdef}).

We note that condition (\ref{NEC1}) ensures that the first term in the integrand \er{integrand} is always non-negative.
Therefore, the conditions
\beq
\label{suff}
f(r) \, g(r)  \geq 0 \qquad \text{and} \qquad A^\prime(r) \geq 0
\eq
will \emph{guarantee} that $dc_d/dr_m \geq 0 $ and hence a monotonic flow for the $c$-function.
While this is a \emph{sufficient} condition for monotonicity, it is certainly not necessary.
A monotonic flow with $c_d^{\,IR} < c_d^{\,UV}$ may still be possible when $f\, g<0$, but in that case
one needs to determine whether the first term in (\ref{integrand})
is strong enough to overcome the second one so that $A^{\prime\prime}(r) \leq 0$ (again assuming $A^\prime \ge 0$).  
Strictly speaking, this is not required for all values of $r$;
as long as the integral in \eqref{genrunning} is non-negative along the entire RG flow 
the $c$-function is monotonic. In this sense, $S_{EE}$ `averages out' any sufficiently small violations of NEC.

Finally, note that
a \emph{necessary} condition for violations of monotonicity is that
the integrand (\ref{integrand}) changes sign
at some \emph{intermediate} location.
However, in order for (\ref{suff}) and conditions of this type 
to be useful one needs to know the entire geometry -- either analytically or numerically.
Since often only the asymptotic behaviors of the solutions are known, we would like to rephrase these conditions
entirely in terms of the UV (or IR) data, when possible.

{\bf UV criteria for monotonicity:}
For simplicity let us start from the special case where $f$ is a constant along the RG flow, so that $f^\prime = 0$.
The integrand (\ref{integrand}) will have a definite sign -- making $c_d$ monotonic --
as long as $g$ does not change sign along the RG trajectory.
Recalling $g'\ge 0$ from (\ref{NEC2}), this will be guaranteed to be the case when $g_{UV} \leq 0$, 
or alternatively when $g_{IR} \geq 0$.
More information is needed, however, to know whether $c_d$ will increase or decrease along the flow.

First, note that having
\beq
\boxed{f = \const \le 0 \quad \text{and} \quad g_{UV} \leq 0 \; }
\eq
is a \emph{sufficient} condition for a monotonic flow with $dc_d/dr_m \geq 0$,
since it ensures that (\ref{suff}) is satisfied.
Second, when
\beq
f = \const \le 0 \quad \text{and} \quad g_{IR} \geq 0
\eq
the candidate $c$-function will still flow monotonically but this time in the `wrong direction,' \emph{increasing}
towards the IR.
This will be the case \emph{e.g.} for a black hole solution of the form
\beq
ds^2_{d+1} = L^2 \left[- \rho^2 h(\rho) \, dt^2 + \frac{d\rho^2}{\rho^2 h(\rho) }  + \rho^2 d\vec{x}^2 \right]
\eq
for which (after an appropriate change of coordinates) we find $f=-1/L$ and $g_{IR} \propto \rho_h^{d-1} T$,
where $\rho_h$ is the horizon radius and $T$ the temperature.  Note that $g_{IR}$ is guaranteed to be positive (or zero for the extremal $T=0$ case).
The 4D Reissner-Nordstrom AdS black hole falls into this category, with
$h(\rho)= 1+\frac{\mu^2}{\rho^4}-\frac{1+\mu^2}{\rho^3}$ and $\mu$ the chemical potential.
Similar behaviors can be seen in the analytical black hole solutions found in \cite{Gubser:2009qt}.
In that case $f^\prime >0$ but $f\,g <0$, and the second term in \eqref{integrand} always dominates over the first, 
forcing (\ref{integrand}) to be negative along the RG flow (with also $A^\prime >0$).
Once again we find a monotonic flow with the wrong sign.  This naive increase in the degrees of freedom towards 
the IR may simply be an indication that $S_{EE}$ approaches the thermal entropy for large $\ell$.

We now relax the assumption of constant $f$, and go back to considering solutions that are completely generic.
We want to examine the relations (\ref{suff})
which guarantee $d c_d/d r_m \geq 0$, and ask when they can be realized.
In general, the conditions (\ref{NEC1}) and (\ref{NEC2}) do not constrain the signs of $f$ and $g$.
However, if additionally
\be\label{fgnp}
\boxed{f_{UV} \le  0 \quad \text{and} \quad g_{UV} \le 0 \; }
\ee
then the NEC imply that both $f$ and $g$ are non-positive along the entire RG flow.
Thus, whenever (\ref{fgnp}) holds, $f\,g\ge0$ everywhere
and the candidate $c$-function behaves as it should, decreasing monotonically towards lower energies
($A^\prime(r_m) \geq 0$  immediately follows from $f \propto - A^\prime$).
Note that this argument relies only on the UV data and is entirely insensitive to what happens in the IR.
We emphasize once again that while (\ref{fgnp}) is a sufficient condition for monotonicity, it is certainly not necessary.

We now study specific cases in which the geometry may satisfy \eqref{fgnp}.
If we have pure AdS in the UV (but allow for generic Lorentz-violating geometries in the IR), then
$A = B = r/L$ in some finite neighborhood of the UV, giving $f_{UV} = - 1/L$ and $g_{UV}=0$,
which automatically satisfy \eqref{fgnp}.
This can happen in models where Lorentz invariance is broken only by sources
such as matter densities that sit \emph{deep in the bulk} and do not extend
to the boundary, so that beyond such sources the UV geometry is pure AdS.

On the other hand, if the geometry is only asymptotically AdS, subleading corrections to the UV metric
may give non-negligible contributions to $g_{UV}$.  Let us write these corrections as
\ba
\label{AdScorrections}
e^{2A(r)} &= e^{2r/L} \[1 + \gamma_A e^{-p r/L} + \cdots \] \,, \nn \\
e^{2B(r)} &= e^{2r/L} \[1 + \gamma_B  e^{-p r/L} + \cdots \] \,,
\ea
where $\gamma_A$ and $\gamma_B$ are coefficients that encode the leading deviation from pure AdS in the UV.
In other words, we assume that the theory goes to a UV fixed point where Lorentz invariance is preserved,
but it can be broken along the RG flow, with the breaking encoded as $\gamma_A \neq \gamma_B$.
The fall-offs can be determined from the behavior of the equations of motion near the UV boundary.
In particular, in the absence of sources we have $p=d$ and the coefficients
$\gamma_A,$ $\gamma_B$ are the VEVs of components of the dual stress tensor
-- respectively, $\lag T_{ii} \rag$ and $\lag T_{00} \rag$.
Plugging (\ref{AdScorrections}) into the definition of $g$, we find
\be
g(r) = \fr{p}{2L} (\gamma_A-\gamma_B) \, e^{(d-p) r/L} + \cdots \,.
\ee
We first note that if $p>d$ we may safely conclude that $g_{UV}=0$ and we have a monotonic flow.  
If $p\le d$, in order to ensure that $g_{UV}\le 0$ we require
\beq
\label{gcond}
\gamma_A - \gamma_B < 0 \, .
\eq
Note that if $\gamma_A = \gamma_B$ we may need to consider more subleading corrections in \er{AdScorrections}.  
The NEC $g'\ge 0$ combined with \er{gcond} means that $p\ge d$, so we have to consider cases with no source here.
Finally, $f^\prime$ to leading order is
\beq
f^\prime(r) = \frac{p}{2L^2} \left[ (\gamma_A - \gamma_B) - p \gamma_A \right] e^{-p r/L} +\cdots \,,
\eq
telling us that the NEC $f^\prime \geq 0$ requires
\beq
\label{fpcond}
\gamma_A - \gamma_B \geq p \, \gamma_A \, .
\eq
Geometries that approach asymptotically AdS in the UV automatically satisfy conditions \er{cond1} and \er{cond2}.
Thus, if we further satisfy the constraints \er{gcond} and \er{fpcond}, we will be guaranteed
a monotonic flow for the $c$-function.

We should mention that there are models in which the sufficient condition \eqref{fgnp} clearly cannot be satisfied.
For example, if a solution which breaks Lorentz invariance at some energy scale flows to $AdS_{d+1}$
in the IR, then we necessarily have $g_{IR}=0$, from which we deduce that $g\ge0$ along the entire RG flow.
Since at the scales where the solution is Lorentz-violating $g\ne0$,
we must have $g_{UV}>0$, violating \eqref{fgnp}.
However, this does not automatically imply that the $c$-function is not monotonic, since \eqref{fgnp}
is only a sufficient condition.

A similar argument implies that \eqref{fgnp} cannot be satisfied if the solution flows to a Lifshitz fixed point in the IR.
This can be easily seen by introducing an effective Lifshitz parameter $z_{eff}\equiv B^\prime/A^\prime$
as done in \cite{Liu:2012wf}, which reduces to $z$ at the fixed points of the flow.
Our condition \eqref{fgnp} can be shown to imply $z_{eff} \le 1$ along the RG flow.
At a Lifshitz fixed point, however, the NEC forces $z>1$.
Thus \eqref{fgnp} cannot be satisfied.
We expect this conclusion to change if in the IR one allows for hyperscaling violation ($\theta \neq 0)$ in addition to Lifshitz scaling.
In particular, we anticipate that $\{z,\theta\}$ scaling solutions which approach AdS in the UV
can satisfy \eqref{fgnp} for appropriate parameter ranges.
This may be checked explicitly (numerically) by using
\emph{e.g.} the solutions of \cite{Bhattacharya:2012zu,Kundu:2012jn}.

To partially motivate this statement, we conclude by showing that geometries which exhibit both Lifshitz scaling
and hyperscaling violation can indeed obey (\ref{fgnp}).
For these scaling solutions the metric is
\bea
\label{scalingsols}
A(r) &=& \left(1-\frac{d-1}{\theta}\right) \ln \left( \frac{\theta}{d-1}\, r \right) \, , \nn\\
B(r) &=& \left(1-\frac{z(d-1)}{\theta}\right) \ln \left( \frac{\theta}{d-1}\, r \right) \, ,
\ea
from which we find
\ba
f &= \(\fr{d-1}{\theta} -1\) \(\fr{\theta}{d-1} \, r\)^{\fr{(d-1)(z-1)}{\theta}-1} \,, \nn \\
g &= -\fr{(d-1)(z-1)}{\theta} \(\fr{\theta}{d-1} \, r\)^{-\fr{(d-1)(d-1+z-\theta)}{\theta}}.
\ea
Imposing the NEC we recover the usual constraints \cite{Dong:2012se}
\bg
(d-1-\theta)\[(d-1)(z-1)-\theta\] \ge0 \, , \nn \\
(z-1)(d-1+z-\theta) \ge 0 \, .
\label{scalingNEC}
\eg
However, we can require further that $f \leq 0$ and $g \le 0$, which, combined with (\ref{scalingNEC}), become
\bg
\fr{d-1-\theta}{\theta} \le0, \quad
\fr{(d-1)(z-1)-\theta}{\theta} \le0, \nn \\
\fr{z-1}{\theta} \ge0, \quad
\fr{d-1+z-\theta}{\theta} \ge0.
\eg
These conditions -- and therefore (\ref{fgnp}) -- are satisfied if
\bg
1 \le z \le \fr{2d-2}{d-2} \,,\quad\\
(d-1)\max\{1,z-1\} \le \theta \le d-1+z \,,
\eg
where we implicitly assumed $d\ge2$.  Further imposing the conditions \er{cond1} and \er{cond2} gives an additional constraint
\be
d\(1-\fr{d-1}{\theta}\) > 1 \qu \Leftrightarrow \qu \theta>d \,.
\ee

Thus, we have identified a class of geometries which obey the
sufficient condition (\ref{fgnp}), ensuring that the $c$-function decreases monotonically towards the IR.
While one should keep in mind that exact hyperscaling violating solutions are not RG fixed points, 
similar arguments should apply to geometries which exhibit $\{z,\theta\}$ scaling only in some intermediate regime.
Note however that our metric ansatz (\ref{metric}) does not allow us to study the spatially modulated 
instabilities \cite{Cremonini:2012ir,Iizuka:2013ag} expected to arise in the deep IR of some of these scaling solutions.

As we have seen, the candidate $c$-function that we extracted
from entanglement does not always decrease monotonically under RG flows
in quantum field theories that violate Lorentz invariance.
However, in certain cases it is still possible to identify simple criteria on the asymptotic behavior of the
geometry that will ensure monotonicity.
Such criteria can then be valuable \emph{guiding principles} for the construction of monotonic RG trajectories --
especially for cases in which the breaking of Lorentz invariance occurs deep in the bulk, where new non-trivial
phases may emerge.
Conversely, the monotonicity of the $c$-function can also be useful for excluding possible IR behaviors of the theory.

While we have restricted our attention to the entanglement entropy of a strip-shaped region
(which allows for analytical computations),
it would be valuable to extend our arguments to spherical regions as well as more general shapes.
Finally, it would be interesting to explore how our analysis compares to the expectations of \cite{Swingle:2013zla}
for weakly coupled field theories, and to the monotonic flows between Bianchi attractors which were
recently identified in \cite{Kachru:2013voa} (although in a different context).

{\bf Acknowledgements:}
We would like to thank S.~Kachru, J.~T.~Liu, B.~Swingle, P.~Szepietowski and in particular K.~Jensen for valuable discussions.   
S.C. was supported by the Cambridge-Mitchell Collaboration in Theoretical Cosmology,
and the Mitchell Family Foundation.
X.D. was supported by the National Science Foundation under grant PHY-0756174.


\begin{thebibliography}{99}

\bibitem{Zamolodchikov:1986gt}
  A.~B.~Zamolodchikov,
  JETP Lett.\  {\bf 43}, 730 (1986)
  [Pisma Zh.\ Eksp.\ Teor.\ Fiz.\  {\bf 43}, 565 (1986)].

\bibitem{Casini:2004bw}
  H.~Casini and M.~Huerta,
  Phys.\ Lett.\ B {\bf 600}, 142 (2004)
  [hep-th/0405111].

\bibitem{Casini:2006es}
  H.~Casini and M.~Huerta,
  J.\ Phys.\ A {\bf 40}, 7031 (2007)
  [cond-mat/0610375].

\bibitem{Casini:2012ei}
  H.~Casini and M.~Huerta,
  Phys.\ Rev.\ D {\bf 85}, 125016 (2012)
  [arXiv:1202.5650 [hep-th]].

\bibitem{Komargodski:2011vj}
  Z.~Komargodski and A.~Schwimmer,
  JHEP {\bf 1112}, 099 (2011)
  [arXiv:1107.3987 [hep-th]].

\bibitem{Komargodski:2011xv}
  Z.~Komargodski,
  JHEP {\bf 1207}, 069 (2012)
  [arXiv:1112.4538 [hep-th]].

\bibitem{Ryu:2006bv}
  S.~Ryu and T.~Takayanagi,
  Phys.\ Rev.\ Lett.\  {\bf 96}, 181602 (2006)
  [hep-th/0603001].

\bibitem{Ryu:2006ef}
  S.~Ryu and T.~Takayanagi,
  JHEP {\bf 0608}, 045 (2006)
  [hep-th/0605073].

\bibitem{Nishioka:2009un}
  T.~Nishioka, S.~Ryu and T.~Takayanagi,
  J.\ Phys.\ A {\bf 42}, 504008 (2009)
  [arXiv:0905.0932 [hep-th]].

\bibitem{Takayanagi:2012kg}
  T.~Takayanagi,
  Class.\ Quant.\ Grav.\  {\bf 29}, 153001 (2012)
  [arXiv:1204.2450 [gr-qc]].

\bibitem{Swingle:2013zla}
  B.~Swingle,
  arXiv:1307.8117 [cond-mat.stat-mech].

\bibitem{Myers:2012ed}
  R.~C.~Myers and A.~Singh,
  JHEP {\bf 1204}, 122 (2012)
  [arXiv:1202.2068 [hep-th]].

\bibitem{Hoyos:2010at}
  C.~Hoyos and P.~Koroteev,
  Phys.\ Rev.\ D {\bf 82}, 084002 (2010)
  [Erratum-ibid.\ D {\bf 82}, 109905 (2010)]
  [arXiv:1007.1428 [hep-th]].

\bibitem{Liu:2012wf}
  J.~T.~Liu and Z.~Zhao,
  arXiv:1206.1047 [hep-th].


\bibitem{Gubser:2009qt}
  S.~S.~Gubser and F.~D.~Rocha,
  Phys.\ Rev.\ D {\bf 81}, 046001 (2010)
  [arXiv:0911.2898 [hep-th]].


\bibitem{Dong:2012se}
  X.~Dong, S.~Harrison, S.~Kachru, G.~Torroba and H.~Wang,
  JHEP {\bf 1206}, 041 (2012)
  [arXiv:1201.1905 [hep-th]].

\bibitem{Bhattacharya:2012zu}
  J.~Bhattacharya, S.~Cremonini and A.~Sinkovics,
  arXiv:1208.1752 [hep-th].

\bibitem{Kundu:2012jn}
  N.~Kundu, P.~Narayan, N.~Sircar and S.~P.~Trivedi,
  arXiv:1208.2008 [hep-th].

\bibitem{Cremonini:2012ir}
  S.~Cremonini and A.~Sinkovics,
  arXiv:1212.4172 [hep-th].

  \bibitem{Iizuka:2013ag}
  N.~Iizuka and K.~Maeda,
  arXiv:1301.5677 [hep-th].

\bibitem{Kachru:2013voa}
  S.~Kachru, N.~Kundu, A.~Saha, R.~Samanta and S.~P.~Trivedi,
  arXiv:1310.5740 [hep-th].

\end{thebibliography}
\end{document}